# Formation of DY Center as n-type Limiting Defect in Bi-Doped Hybrid Halide Perovskites

*Jin-Ling Li[1,2], Jingxiu Yang[2]\*, Tom Wu[3]\*, and Su-Huai Wei[2]\**

[1] International Center for Quantum Materials, School of Physics, Peking University, Beijing 100871, China

[2] Beijing Computational Science Research Center, Beijing, 100193, China

[3] School of Materials Science and Engineering, University of New South Wales (UNSW), Sydney, NSW 2052, Australia

AUTHOR INFORMATION

**Corresponding Author**

\* Jingxiu Yang, Email: yangjingxiu@csrc.ac.cn

\* Tom Wu, Email: tom.wu@unsw.edu.au

\* Su-Huai Wei, Email: suhuaiwei@csrc.ac.cn

**ABSTRACT**

It is well known that the DX center is a kind of defect that limits the n-type doping in some tetrahedral coordinated semiconductors. It is a deep negatively charged defect complex converted from a nominal shallow donor defect, which can serve as a trap center of electrons, thus is detrimental to the performance of optoelectronic devices. Similar to the DX center, we find that a donor-yielded complex center (DY center) also exists in six-fold coordinated semiconducting materials. For example, Bi is commonly expected as a shallow n-type dopant in perovskite $APbX_3$. However, our first-principles calculations show that the DY center is formed in Bi-doped $MAPbBr_3$ when the Fermi level is high in the gap, but, interestingly, it does not form in $MAPbI_3$. The reason that the DY center is formed in $MAPbBr_3$ instead of $MAPbI_3$ is attributed to the high conduction band minimum (CBM) of $MAPbBr_3$. Our results are able to explain recent puzzling experiment observations and the thorough discussions of the formation and the properties of the DY center in perovskites provide enlightening insights to the defect study in six-fold coordinated semiconductors.

**TOC GRAPHICS**

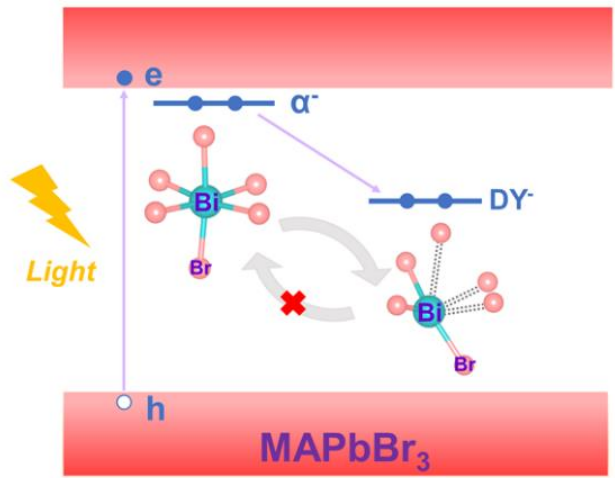

# 1. INTRODUCTION

Donor-complex center (DX center) is often observed in four-fold coordinated zincblende (ZB) or wurtzite (WZ) III-V[1] and II-VI[2,3] semiconductors which limits n-type doping. For example, Si substituting cation in (Al,Ga)As alloy can create either shallow substitutional donor $Si_{Ga}^+$ or a negatively charged defect complex center DX$^-$ when the Fermi energy approaches the conduction band minimum (CBM), which acts as a limiting defect for n-type doping in the system. This special DX$^-$ defect center is often associated with large atomic displacements and the transition energy levels of the DX center is usually deep inside the band gap with localized wavefunctions. The formation of DX center is extremely unfavorable for optoelectronic devices, so it has been extensively studied for tetrahedron semiconductors. However, the report of DX-like center is rare in six-coordinated semiconductors (to distinguish it from the DX center in the tetrahedral system, we will denote it as DY center). It is not clear what the local geometry of the DY center and what their electronic structures are in the octahedral system.

MAPbX$_3$ (MA=CH$_3$NH$_3^+$, X=I$^-$ or Br$^-$), have attracted tremendous attentions in the field of photovoltaic and photoluminescent materials owing to their excellent optoelectronic properties, such as strong optical absorption and emission in the visible range, long carrier lifetime and diffusion length, shallow intrinsic defects and low nonradiative recombination rates.[4-9] Controlled by their respective dominant intrinsic defects, MAPbI$_3$ was reported to exhibit both p-type and weak n-type ambipolar conductivities, whereas MAPbBr$_3$ exhibits mainly p-type conductivity.[10-12] To enhance the electron conductivity, Bi is used as an n-type dopant in MAPbI$_3$ and MAPbBr$_3$ in experiment, because Bi as a neighboring element to Pb in the Periodic Table is expected to act as a shallow donor when it substitutes on Pb site, forming Bi$_{Pb}$.[13,14] However, it was found experimentally that the photocarrier lifetime was shorter and the photoluminescence (PL) was suppressed in Bi-doped MAPbBr$_3$. These phenomena were attributed to the enhanced carrier trapping, suggesting that Bi dopants might form deep defects in MAPbBr$_3$.[15,16] However, the exact type of the defect is unknown.

In this work, we systematically investigate the substitutional defect structures and properties of Bi$_{Pb}$ in MAPbBr$_3$ and MAPbI$_3$ based on the first-principles calculations. We find that Bi prefers to be a donor in MAPbI$_3$ but might induce DY center in MAPbBr$_3$, which converts a shallow donor defect to a deep acceptor defect center in MAPbBr$_3$. The DY center can trap electrons and therefore decrease the carrier lifetime and suppress the PL. The local atomic structure of the DY center in the octahedral environment is revealed. Our results suggest that similar to DX center in tetrahedral semiconductor, DY defect center can also form in octahedral semiconductors that can limit n-type doping in the system. This can have large impact on the device applications of octahedron semiconductors.

## 2. COMPUTATIONAL METHODS

The first-principles calculations are based on the density functional theory (DFT) as implemented in the VASP code.[17] The projector augmented-wave (PAW)[18,19] pseudopotentials and the general gradient approximation (GGA) as parameterized by Perdew-Burke-Ernzerhof for solids (PBEsol)[20] are employed. The cutoff energy for the plane wave basis is 400 eV. $5d^{10}6s^26p^2$ and $5d^{10}6s^26p^3$ valence electrons are adopted for Pb and Bi, respectively, in all the calculations. The 3×3×3 supercell combined with a 4×4×4 Monkhorst-Pack sampling is employed for the defect calculations. All atoms are allowed to relax until the forces on atoms are below 0.02 eV/Å. The MA molecule oriented along <100> direction is found to be the most stable orientation among all the high symmetric directions according to PBEsol functional calculations.[21] The calculated lattice constant and band gap are 5.84 Å and 1.88 eV for the pseudocubic MAPbBr$_3$, which is 0.1 Å and 0.47 eV smaller than that of the experiment values, respectively.[22,23] The calculated lattice constant and band gap are 6.23 Å and 1.51 eV for the pseudocubic MAPbI$_3$, which are in good agreement with the experimental and previous calculations.[23-25]

The calculations for the defect formation energies and transition energy levels follow the established methods.[26] The formation energy is defined as:

$$\Delta H_f(q) = \Delta E(q) + \Sigma n_i \mu_i + qE_F \quad (1)$$

Where $\Delta E(q) = E(q) - E(\text{host}) + \Sigma n_i E(i) + q\varepsilon_{VBM}(\text{host})$. $E_F$ is the Fermi energy and referenced to the valence band maximum (VBM) of the host. $\mu_i$ is the chemical potential of component element i referenced to E(i) for the most stable phase. To avoid the formation of the possible secondary phases, the following constraints of the atomic chemical potential should be satisfied:

$$\mu_{MA} + \mu_{Pb} + 3\mu_{Br} = \Delta H_f(MAPbBr_3) = -6.19 \text{ eV} \quad (2)$$

$$\mu_{MA} + \mu_{Br} < \Delta H_f(MABr) = -3.52 \text{ eV} \quad (3)$$

$$\mu_{Pb} + 2\mu_{Br} < \Delta H_f(PbBr_2) = -2.64 \text{ eV} \quad (4)$$

In our calculations, the chemical potential range for the stable MAPbBr$_3$ is narrow, which is consistent with the early studies.[12] For the extrinsic Bi dopant, an additional constraint to exclude the formation of BiBr$_3$ also needs to be met,

$$\mu_{Bi} + 3\mu_{Br} < \Delta H_f(BiBr_3) = -2.92 \text{ eV} \quad (5)$$

Considering all the constraints above, we adopted a chemical potential condition: Br-rich/Pb-poor ($\mu_{MA}$ = -3.52 eV, $\mu_{Pb}$ = -2.67 eV, $\mu_{Br}$ = 0.00 eV, $\mu_{Bi}$ = -2.92 eV) for MAPbBr$_3$ in this work. Similar treatment is also applied for MAPbI$_3$, and the chemical potential $\mu_{MA}$ = -2.91 eV, $\mu_{Pb}$ = -1.82 eV, $\mu_I$ = 0.00 eV, $\mu_{Bi}$ = -1.78 eV is used for I-rich/Pb-poor condition. Note that the formation of the DY center is independent of the atomic chemical potentials.

## 3. RESULTS AND DISCUSSION

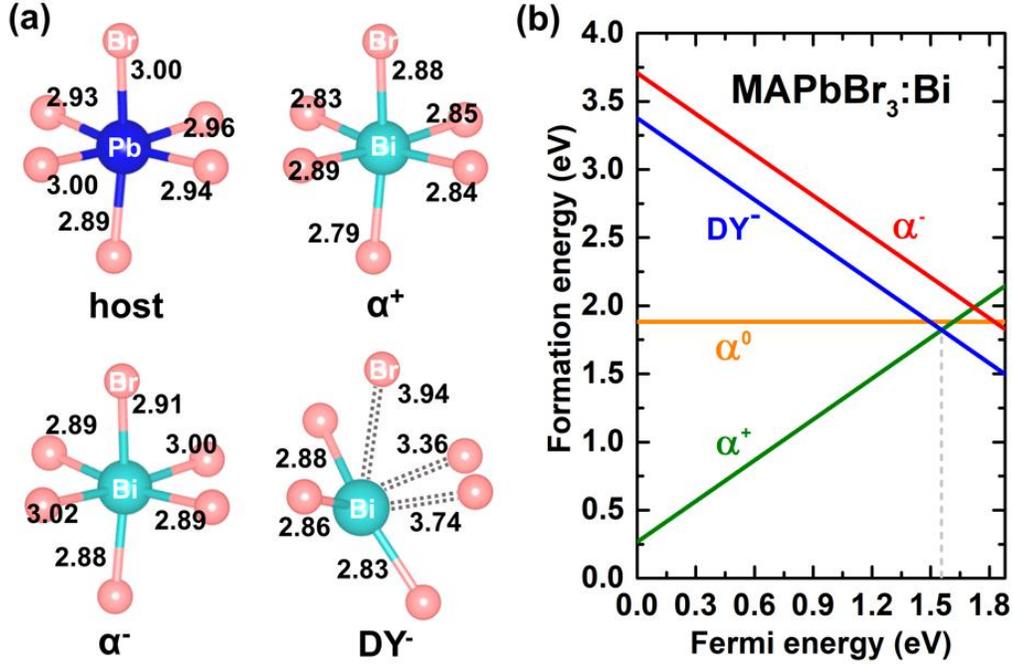

**Figure 1.** (a) The local structures of the host and the $\alpha^+$, $\alpha^-$ and $DY^-$ defect states in Bi-doped MAPbBr$_3$. The numbers indicate the obtained bond lengths for the full lines or distances for the dotted lines of the broken bonds. (b) The calculated defect formation energy of Bi$_{Pb}$ in MAPbBr$_3$ at different charge states as a function of Fermi energy ($E_F$) under the Br-rich/Pb-poor condition.

The structure of MAPbBr$_3$ is composed of the PbBr$_6$ octahedron with the MA molecule occupying the A-site. Influenced by the anisotropic MA$^+$ cation, the PbBr$_6$ octahedron is distorted with the average bond length 2.95 Å in the pure MAPbBr$_3$, as shown in Fig. 1(a). For Bi-doped MAPbBr$_3$, the possible charge states of $\alpha$(Bi$_{Pb}$) include +1, 0 and -1. For the positive charged $\alpha^+$ state, the Bi$^{3+}$ ion has the same electron configuration as Pb$^{2+}$ ion, so the local structure of $\alpha^+$ state is still six-fold coordinated, similar to the PbBr$_6$ octahedron in MAPbBr$_3$. Besides, due to the small ionic radius of Bi$^{3+}$, the average bond length of Bi-Br is shortened to 2.85 Å. For the neutral $\alpha^0$ state, the local structure is still six-fold coordinated with the average bond length of 2.91 Å. When forming the negatively charged state, two possible structures are found, which are the normal $\alpha^-$ state and the highly distorted DY complex center (DY$^-$) state, as shown in Fig. 1(a), bottom panel. With two redundant electrons, the $\alpha^-$ state is six-fold coordinated with the average bond-length 2.93 Å, which is further elongated compared to those of $\alpha^+$ and $\alpha^0$ state due to the Coulomb repulsion. For the negatively charged DY$^-$ state, the central Bi atom is highly distorted moving towards the <111> direction, hence changing the local structure from six-fold to three-fold coordination by breaking three Bi-Br bonds. The residual three bond lengths are not exactly the same due to the existence of the MA molecule. To determine the stability of the DY$^-$ state, we calculated the DY formation energy ($\Delta E(DY)$), which is defined as[27]

$$\Delta E(DY) = E(DY^-) - E(\alpha^-) \quad (6)$$

Here $E(DY^-)$ and $E(\alpha^-)$ are the total energies of the DY$^-$ and the $\alpha^-$ defects, respectively. A negative value of $\Delta E(DY)$ would indicate that the DY$^-$ state is more stable than the $\alpha^-$ state. The DY formation energy is calculated to be -0.28 eV for MAPbBr$_3$. It is possible to form DY center in MAPbBr$_3$, from the energy standpoint.

The calculated formation energy of Bi$_{Pb}$ defect in MAPbBr$_3$ as function of Fermi energy is shown in Fig. 1(b). The transition energy $\varepsilon(0/+)$ of Bi$_{Pb}$ in MAPbBr$_3$ is 0.26 eV below the CBM and the transition energy $\varepsilon(0/-)$ is 0.38 eV below the CBM for the DY$^-$ state, so it is an negative U system, i.e., the neutral charged state $\alpha^0$ is unstable with respect to the dissociation into DY$^-$ state and $\alpha^+$ state. This negative U behavior is a typical character of the defect state with large atomic relaxations.[17] In this specific case, when the Fermi energy is below the Fermi energy corresponding to the crossing point, that is 0.32 eV below the CBM, the positive $\alpha^+$ is dominant. On the other hand, if the Fermi energy exceeds the crossing point, it prefers to form the localized DY⁻ state. Therefore, the Fermi energy will be pinned at about 1.55 eV with comparable α⁺ and DY⁻ states.

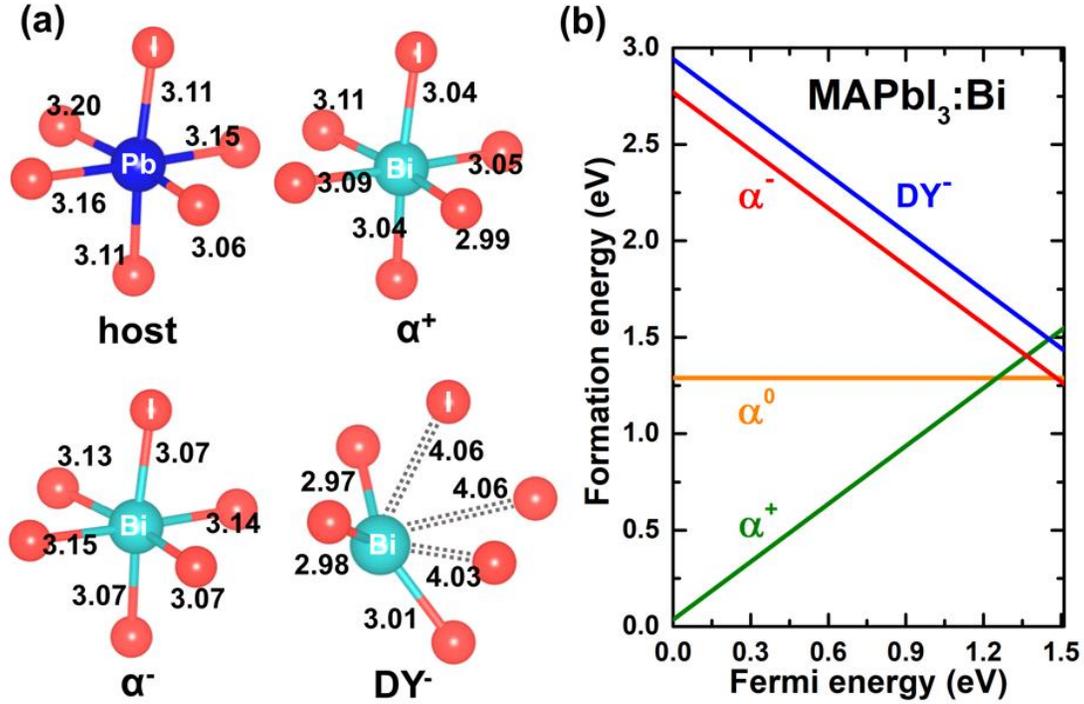

**Figure 2.** (a) The local structures of the host and the α⁺, α⁻ and DY⁻ defect states in Bi-doped MAPbI$_3$. The numbers indicate the obtained bond lengths for the full lines or distances for the dotted lines of the broken bonds. (b) The calculated defect formation energy of Bi$_{Pb}$ in MAPbI$_3$ at different charge states as a function of Fermi energy ($E_F$) under the I-rich/Pb-poor condition.

The structures of host and Bi$_{Pb}$ defect states in MAPbI$_3$ are all six-coordinated, as shown in Fig. 2(a), and the corresponding average Bi-I bond lengths are 3.05, 3.09 and 3.11 Å, respectively, for the α⁺, α⁰ and α⁻ states compared to 3.13 Å for the host Pb-I bond. For the DY⁻ state, the geometry is similar to that in MAPbBr$_3$. The formation energy of the DY⁻ state is calculated to be 0.14 eV, indicating that the octahedral α⁻ state is thermodynamically more stable in MAPbI$_3$. Therefore, the stable Bi$_{Pb}$ defects would remain octahedral coordination in MAPbI$_3$.

The calculated formation energies of the $Bi_{Pb}$ defect at different charge states are shown in Fig. 2(b). The transition energy $\varepsilon(0/+)$ and $\varepsilon(0/-)$ of $Bi_{Pb}$ in $MAPbI_3$ is calculated to be 0.26 eV and 0.03 eV below the CBM, respectively. When the Fermi energy is below 1.25 eV (-0.26 eV vs. the CBM), the stable state of the $Bi_{Pb}$ defect is $\alpha^+$ state. As the Fermi level increases beyond 1.25 eV, the stable state converts from $\alpha^+$ to $\alpha^0$ state. When the Fermi level reaches 1.48 eV (-0.03 eV vs. the CBM), the stable state is then converted to $\alpha^-$ state. Generally speaking, the Bi dopant serves as a donor in $MAPbI_3$ as expected and the system exhibit a positive U behavior, i.e., two neutral states are more stable than the dissociative plus and minus charged states.

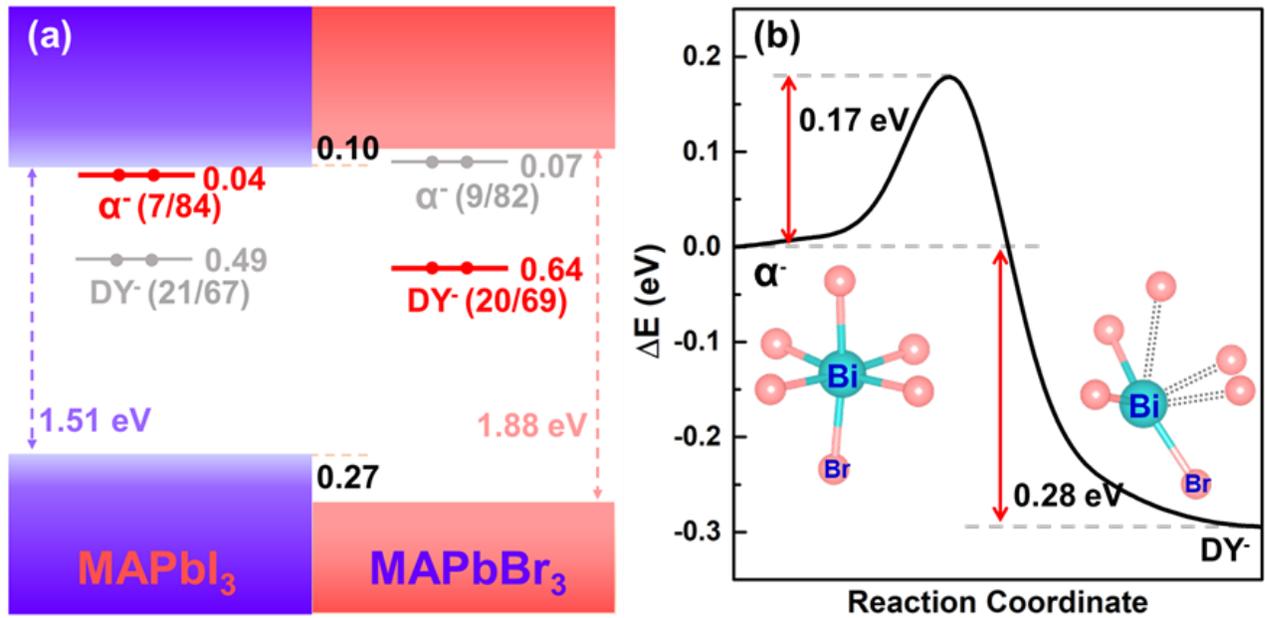

**Figure 3.** (a) The alignment of the band edges of $MAPbI_3$ and $MAPbBr_3$ and the relative single electron energy levels of $\alpha^-$ and $DY^-$ states. The red lines represent the stable states, while the grey lines represent the metastable states. The numbers in bracket represent the projected *p* orbitals of (anion/cation). (b) The transition path for Bi-doped $MAPbBr_3$ from the metastable $\alpha^-$ state to the stable $DY^-$ state.

As mentioned above, the properties of the $Bi_{Pb}^+$ states in MAPbI$_3$ and MAPbBr$_3$ are quite similar because the hosts share the common electronic structures except for the band gap. However, the properties of the $Bi_{Pb}^-$ state are quite different in MAPbI$_3$ and MAPbBr$_3$. To understand the difference, we have further analyzed the components of the defect. The VBMs of the perovskites are predominantly composed by the anion valence *p* orbitals. The CBMs are mainly contributed by the Pb 6*p* orbitals. The defect levels of the α⁻ states in MAPbI$_3$ and MAPbBr$_3$ are predominantly contributed by anion valence *p* orbitals (see Fig. S1 and Fig. S3 in Supplementary Information, SI), similar to the CBMs, and their energy follow that of the respective CBM as one would expect for shallow defects, as shown in Fig. 3(a). In contrast, the localized DY⁻ states in the two perovskites are contributed by the comparable anion valence *p* and cation Bi 6*p* orbitals (see Fig. S2 and Fig. S4 in SI) and their defect levels do not change much as the CBMs or the VBMs shift in the two systems. Therefore, the electronic energy gain by moving the two electrons from the α⁻ state to the DY⁻ state in MAPbBr$_3$ is higher than that in MAPbI$_3$ because of the higher CBM of MAPbBr$_3$ than MAPbI$_3$. This explains why DY⁻ state in MAPbBr$_3$ is stable but it is not in MAPbI$_3$, and why, in general, the DY⁻ state is more stable when the CBM is higher. Note that the CBM of MAPbBr$_3$ should be actually higher because of the underestimate of its band gap by the PBEsol functional calculation, we expect the prediction of formation of the DY⁻ center in MAPbBr$_3$ is robust.

Used in the optoelectronic or photoluminecsent devices, Bi doped MAPbBr$_3$ will inevitably be excited by external illumination. When the light is off, the shallow n-type α⁺ state is dominant, which could result in n-type electric conductance as found in the experiment.[14] Under light illumination, the photo-exited electrons would raise the quasi-Fermi energy to be close to the CBM. Based on our results, the α⁺ state is expected to convert to the α⁻ state at first due to the little distortion accepting the additional electrons. However, the α⁻ state is metastable and could easily transform to the DY⁻ state after overcoming the small energy barrier of 0.17 eV, as shown in Fig. 3(b). On the other hand, the DY⁻ state could hardly convert back to the α⁻ state due to the 0.45 eV energy barrier backwards. The localized DY⁻ state is deep and could trap the photo-electrons, which might cause the suppressed PL and decreased photocurrent as suggested by the experiments.[15,16]

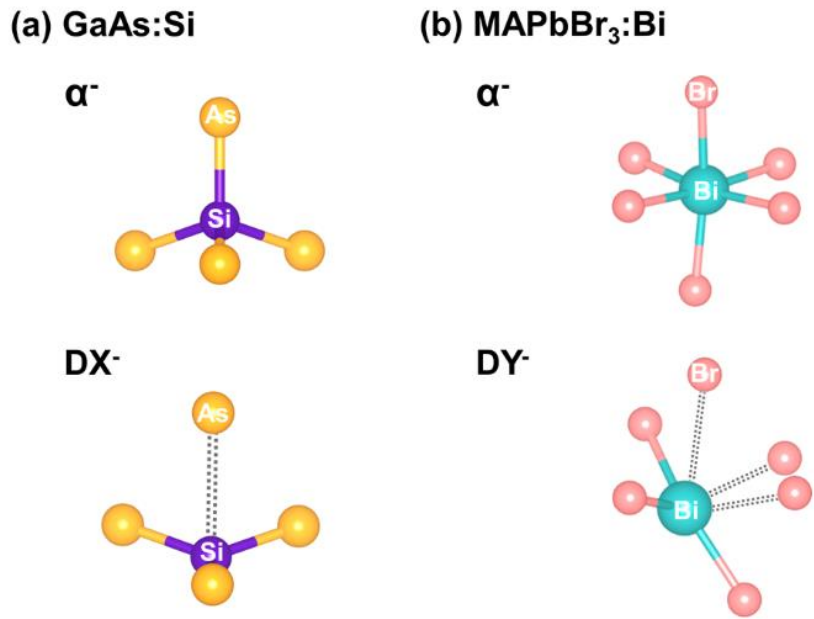

**Figure 4.** The comparison of the α⁻ and the DX⁻ (DY⁻) states in tetrahedral (a) and octahedral (b) coordination.

To compare the DX⁻ and the DY⁻ states in four- and six-coordinated compounds, the $Si_{Ga}^-$ defect in GaAs and $Bi_{Pb}^-$ defect in MAPbBr$_3$ are shown as examples in Fig. 4. The DY⁻ and DX⁻ state are both highly distorted structures with localized deep transition levels. The corresponding tetrahedral and octahedral α⁻ states are shallow so that the defect formation energy increases as the CBM rises. However, the DX⁻ and DY⁻ states are deep, thus the defect formation energy is hardly affected by the position of the CBM. As a result, as the CBM rises, the DX and DY formation energies are reduced or become more negative in four- and six-fold coordinated compounds. The difference between the DX$^-$ and DY$^-$ states is about their local structures. As shown in Fig. 4(a), the DX$^-$ state is formed by breaking a single bond, hence changing the local symmetry from T$_d$ to C$_{3v}$. On the other hand, the DY$^-$ center in Fig. 4(b) is formed by breaking three bonds, changing the local symmetry from O$_h$ to C$_{3v}$, if the influence of MA molecule is ignored. This difference is caused by the different local environment in the two systems.

## 4. CONCLUSIONS

Based on the first-principles calculations, we have identified a DY center in Bi doped MAPbBr$_3$, which is a deep localized trapping state, limiting the n-type doping in the system and could cause negative photoconductivity. However, in spite of the similar electronic structure of MAPbBr$_3$ and MAPbI$_3$, the DY center is not formed in MAPbI$_3$. We show that the higher CBM in MAPbBr$_3$ is the main reason led to the negative formation energy of the DY center in MAPbBr$_3$. The DY states in MAPbBr$_3$ are highly localized and deep enough to serve as the trap centers for electrons or non-radiative recombination centers. Hence, doping Bi into MAPbI$_3$ and MAPbBr$_3$ may result in different performance in photovoltaics and photoluminescence under light radiation. The discussion of DY center and comparison with related DX center in four-fold coordinated system provided deep insights and guideline for future study of the defect properties in four- and six-fold coordinated semiconductors.

**AUTHOR INFORMATION**

**Corresponding Author**

* Jingxiu Yang, Email: yangjingxiu@csrc.ac.cn

* Tom Wu, Email: tom.wu@unsw.edu.au

* Su-Huai Wei, Email: suhuaiwei@csrc.ac.cn

**Notes**

The authors declare no competing financial interests.

# ACKNOWLEDGMENTS

This work is funded by the National Key Research and Development Program of China under Grant No. 2016YFB0700700 and the Natural Science Foundation of China under Grant No. 51672023; 11634003; and U1530401. We also acknowledge the computer support of TH2-JK at the Beijing Computational Science Research Center (CSRC).